\documentclass[aps,twocolumn]{revtex4}
\usepackage{epsfig}
\newcommand{\be}{\begin{equation}}
\newcommand{\ee}{\end{equation}}
\newcommand{\bea}{\begin{eqnarray}}
\newcommand{\eea}{\end{eqnarray}}
\newcommand{\ba}{\begin{array}}
\newcommand{\ea}{\end{array}}

\begin{document}

\title{Collapsing lattice animals and lattice trees in two dimensions}
\author{Hsiao-Ping Hsu and Peter Grassberger}
\affiliation{John-von-Neumann Institute for Computing, Forschungszentrum
J\"ulich, D-52425 J\"ulich, Germany}
                                                                                
\date{\today}
\begin{abstract}
We present high statistics simulations of weighted lattice bond animals and 
lattice trees on the square lattice, with fugacities for each non-bonded contact and for each bond 
between two neighbouring monomers. The simulations are performed using a newly 
developed sequential sampling method with resampling, very similar to the 
pruned-enriched Rosenbluth method (PERM) used for linear chain polymers. We 
determine with high precision the line of second order transitions from an extended 
to a collapsed phase in the resulting 2-dimensional phase diagram. This line 
includes critical bond percolation as a multicritical point, and we verify  
that this point divides the line into different universality classes. 
One of them corresponds to the collapse driven by contacts and includes the 
collapse of (weakly embeddable) trees. 
There is some evidence that the other is subdivided again into two parts with different
universality classes. One of these (at the far side from collapsing trees)
is bond driven and is represented by the Derrida-Herrmann model of animals having 
bonds only (no contacts). Between the critical percolation point and this bond driven
collapse seems to be an intermediate regime, whose other end point is a 
multicritical point $P^*$ where a transition line between two collapsed phases 
(one bond-driven and the other contact-driven) sparks off. This point $P^*$ seems to 
be attractive (in RG sense) from the side of the intermediate regime, so there are 
four universality classes on the transition
line (collapsing trees, critical percolation, intermediate regime, and Derrida-Herrmann).
We obtain very precise estimates for all critical exponents for collapsing trees. 
It is already harder to estimate the critical exponents for the intermediate regime. 
Finally, it is very difficult to obtain with our method good estimates of the 
critical parameters of the Derrida-Herrmann universality class. As regards the 
bond-driven to contact-driven transition in the collapsed phase, we have some
evidence for its existence and rough location, but no precise estimates of critical
exponents.
\end{abstract}

\maketitle

\section{Introduction}

Lattice animals are just clusters of connected sites on a regular lattice.
They play an important role in many models of statistical physics, as e.g. 
percolation~\cite{Stauffer92} and the Ising model, where they are related 
to Fortuin-Kastleyn clusters and to the famous Swendsen-Wang 
algorithm~\cite{Fortuin,Swendsen}. Apart from this, they form also the prototype 
model for randomly branched polymers, just as self avoiding random walks
(SAWs) are a model for unbranched polymers~\cite{Lubensky}. 

In the case of single unbranched polymers in a very diluted solvent, a much 
studied phase transition happens when the solvent deteriorates (as it usually
does when temperature is lowered). Below the so-called $\theta$-point a polymer 
no longer forms a swollen coil with Flory radius $R_N \sim N^\nu$ with 
$\nu>1/2$ ($N$ is here and in the following the number of monomers), but rather 
a collapsed ``globule" with $R_N \sim N^{1/d}$, where $d$ is the dimensionality
of space~\cite{deGennes,Grosberg}. In the simplest model of polymers living 
on a regular lattice, the collapse is induced by an effective attractive 
interaction between non-bonded monomers on neighbouring lattice sites. It is 
easy to see that such a monomer-monomer attraction is equivalent to a 
monomer-solvent repulsion, so there is no need to include the latter, if one 
is only interested in universal properties of the transition. 

A similar collapse transition is expected to occur also for branched 
polymers. Here the situation is somewhat more complicated, though. In a 
lattice model, one can introduce two different kinds of attraction, so one 
obtains a two dimensional phase diagram with a line of collapse transition 
points. The basic reason for this difference to unbranched polymers 
(interacting self avoiding walks, ISAW) is that the number of bonds is fixed
to $b=N-1$ for SAWs (as it is also for trees), while it can fluctuate, $b\geq N-1$,
for general animals. Thus one can introduce two different fugacities for 
monomer-monomer bonds and for non-bonded monomer-monomer contacts.
As for unbranched polymers, there is no need for introducing a separate
monomer-solvent interaction, since the number $s$ of monomer-solvent contacts
is not independent, but is given by
\be
   {\cal N}N = 2b + 2k + s,                                        \label{1}
\ee
where ${\cal N}$ is the lattice coordination number (${\cal N} = 2d$ on a simple 
hypercubic lattice in $d$ dimensions) and $k$ is the number of non-bonded 
monomer-monomer contacts. A general partition sum for interacting lattice 
animals is therefore~\cite{Henkel96,Seno94,Flesia94,Flesia92,Stratychuk95,Madras97,Rensburg97,Rensburg99,Rensburg00}
\be
   Z_N(y,\tau) = \sum_{b,k} C_{Nbk}\; y^{b-N+1} \tau^k,                    \label{2}
\ee
where $C_{Nbk}$ is just the number of configurations (up to translations 
and rotations) of connected clusters with $N$ sites, $b$ bonds, and $k$ 
contacts. Notice that we changed the definition slightly with 
respect to~\cite{Henkel96,Seno94,Flesia94,Flesia92,Stratychuk95,Madras97,Rensburg97,Rensburg99,Rensburg00}, 
so that $Z_N(y,\tau)$ is non-zero for $y=0$.

The phase diagram in terms of the control parameters $y$ and $\tau$ is shown 
in Fig.~1 for 2-d animals on the square lattice. There is an extended phase 
for small $y$ and $\tau$ which includes 
also the unweighted animal model ($y=\tau=1$), and at least one collapsed 
phase. At the collapse transition we expect that 
\be
   Z_N(y,\tau=\tau_c(y)) \sim \mu(y)^N N^{-\theta}
\ee
where $\mu(y)$ should depend continuously on $y$, but $\theta$ should take 
discrete values depending on the respective universality class.

There have been claims, based on exact enumerations of very small 
clusters~\cite{Flesia92,Flesia94}, that there are two distinct collapsed phases, 
one bond-rich and the other contact-rich. This has led to some controversy, 
since later authors could not find it either with other 
numerical models~\cite{Seno94} or in simplified models~\cite{Henkel96}. 
We also find such a transition between two collapsed phases, although
at significantly smaller values of $y$ (short dashed-dotted curve).
The two end points of the transition curve shown in Fig.~1 are given by 
collapsing weakly embeddable trees ($y=0$) and by a model where non-bonded 
contacts are forbidden ($\tau=0$). The latter has been studied in detail by
Derrida and Herrmann~\cite{Derrida-H}, and others~\cite{Dickman,Lam87,Lam88,Shapir}.

\begin{figure}
  \begin{center}
    \psfig{file=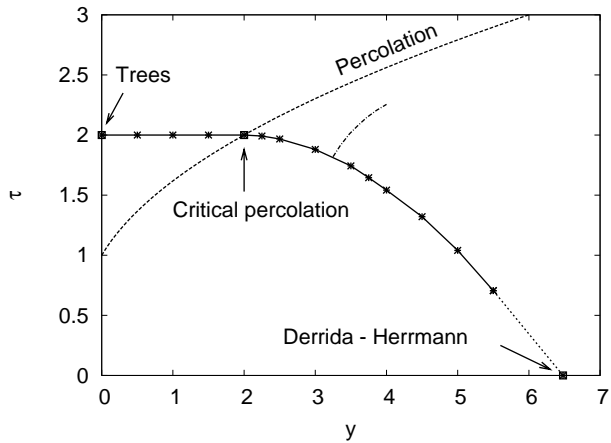,width=6.0cm,angle=270}
    \caption{Phase diagram for interacting animals. The full line separates an extended 
      phase (below) from a collapsed phase (above). The points on this line are obtained 
      by Monte Carlo simulations in the present work, except for the point at $y=6.48, \tau=0$
      which is taken from~\cite{Derrida-H}. At $y=0$ the clusters are trees (minimal number 
      of bonds), while at $\tau=0$ they have no contacts but only bonds. The dashed line 
      corresponds to bond percolation, with the critical point being at $y=\tau=2$. The short 
      dashed-dotted line is a rough estimate for the transition between a contact-rich and a 
      bond-rich collapsed phase.}
\label{fig1}
\end{center}
\end{figure}

Bond percolation is a special model of weighted clusters. The generating 
function for bond percolation clusters is defined as 
\be
   Z^{\rm perc}_N(p) = \sum_{b,k} C_{Nbk}\; p^b (1-p)^{k+s} .                    \label{3}
\ee
More precisely, this equation gives the probability that the origin is 
connected to a cluster of precisely $N$ sites, if lattice bonds are established 
with probability $p$ and broken with probability $1-p$. Using Eq.~(\ref{1}), 
one sees that 
\be
   Z^{\rm perc}_N(p) = (1-p)^{{\cal N}N-2N+2} p^{N-1} Z_N(y(p),\tau(p)),
\ee
with the curve $(y(p),\tau(p))$ parametrized as 
\be
   y=p/(1-p)^2\;,\quad \tau=1/(1-p)\;,\qquad 0\leq p \leq 1 \;.     \label{4}
\ee
or given explicitly by 
\be
   y=\tau(\tau-1)\;.
\ee

This curve is also shown in Fig.~1. It contains in particular the critical 
bond percolation point, $p=1/2$, corresponding to $y=\tau=2$. This point 
lies also on the collapse transition line~\cite{Henkel96}, and divides it 
into two parts: One branch with high density of contacts ($y<2$) and one 
with high density of bonds ($y>2$). Since it is known that the critical 
percolation point is fully repulsive in the renormalization group (RG) 
sense~\cite{nienhuis}, it would be natural to assume that the RG 
flow along the transition curve goes to the two end points $y=0$ (collapsing
trees) resp. $\tau=0$ (Derrida-Herrmann model). Thus we could naively expect 
that the critical behaviour on each of the two parts of the transition curve is 
given by the fixed point at its end. But if there are indeed two different 
collapsed phases, then the point at which the transition line between them meets 
the collapse transition line should be some multicritical point $(y^*,\tau^*)$. 
A priori, this point could be the critical percolation point. 
But according to~\cite{Flesia92,Flesia94,Peard95} this is clearly not the case: while $y=2$ for 
percolation, $y^* \approx 4$ (see Fig.~2 of~\cite{Henkel96}). If all that is true, 
then there should be two
multicritical points on the transition curve, and three different universality
classes in addition to the percolation class.

In the following we shall investigate the details of this scenario by means 
of extensive Monte Carlo simulations. In Sec.~2
we shall shortly describe the algorithm. The critical percolation point is 
studied in Sec.~3, both as a means to verify the exactness of our codes and 
to discuss the cross-over exponents at this point. We shall see that both 
cross-over exponents (the one associated to fluctuations of the number of bonds 
and the other one for contacts) are exactly equal to 1/2, although the 
fluctuations of the number of contacts has huge corrections to scaling.
After that we turn in Sec.~4 to the collapse of trees, i.e. to the case $y<2$. We find that 
there our results are very clean, allowing us to obtain conjectures for the 
exact values of all critical exponents. We also verify with high precision 
a previous conjecture that the transition curve for $y<2$ is strictly 
horizontal, with $\tau = 2$. Next, we will turn to the case $y>2$ in Sec.~5.
There we have much more problems in arriving at a consistent 
scenario, presumably because there exist indeed two different collapsed phases
which meet on the transition curve at a point different from critical percolation.
In particular, at face value our results would suggest that the cross-over exponent 
for bonds depends continuously on $y$, and would thus be not universal. But apart from 
these problems it seems very clear that the critical behaviour for $y>2$ is 
not in the same universality class as that for $y<2$, in contrast to recent
speculations~\cite{Rensburg99}. We follow up the question of a transition between 
two collapsed phases by simulating deep in the collapsed phase.
We conclude with a discussion in Sec.~7.

\section{Numerical Methods}

We use essentially the same method as in~\cite{animals04}. There we introduced a 
sequential sampling method with re-sampling, implemented depth-first as in the 
PERM (pruned-enriched Rosenbluth method) algorithm of Ref.~\cite{PERM}. More 
precisely, we first choose a value of $p$ near the percolation threshold 
(i.e., $p\approx 1/2$), and simulate bond percolation clusters at this $p$ by
means of a variant of the Leath algorithm~\cite{Leath}. While they are still 
growing we estimate their current contribution to the percolation partition 
sum, and -- by re-weighting them according to the animal ensemble, Eq.~(\ref{2}) --
to the animal ensemble. If the latter happens to be higher that average, we 
clone the cluster and let both copies evolve independently further. If the 
weight is too small, we prune with probability 1/2 and increase the weights
of the survivors by a factor 2.

For this to be efficient, we have to take the following considerations into
account:\\
-- The clusters are grown breadth first, not depth first~\cite{animals04},
although the ``population control" is done depth first.\\
-- The optimal value of $p$ depends on $y$ and $\tau$, and has to be found 
by trial and error.\\
-- The optimal value of the pruning/cloning threshold was found, to very 
good precision, to be the same as for ordinary animals~\cite{animals04},
although we could improve also this by trial and error search.\\
-- To make things even more subtle, the optimal parameters depended slightly
on the maximal sizes to be simulated.

To monitor the performance of the algorithm, we used mostly the variance of 
the estimated partition sum. Notice that estimating the partition sum is an
integral part of this algorithm (it is needed for the ``population control").
In addition, we checked that the distributions of ``tour weights" remain
acceptable, as described in~\cite{animals04}.

It is clear that this algorithm should work best near the percolation point
(since there is then a minimal amount of re-sampling needed). Indeed, 
exactly at the percolation point the method works even better than Leath
itself, if one uses $p$ slightly below 1/2. The reason is that with
$p=1/2$ and without re-sampling the sample contains very few large but 
finite clusters. The partition sum Eq.~(\ref{3}) decreases at $p=p_c=1/2$
very slowly with $N$, like $N^{-0.054945}$~\cite{Stauffer92}. Thus most 
clusters which did not stop growing early will do so only very late, beyond
the maximal $N$ we can simulate. Therefore, estimates of specific heat or 
of gyration radii will be based on rather small samples and will have rather 
large errors. With resampling we can work at $p<p_c$, and we will have 
many more large finished clusters in the sample.

All the work reported in this paper was done on fast Linux PCs and used 
nearly 2 years of CPU time.

\section{The Percolation Point}

We use the above algorithm to simulate critical bond percolation clusters 
($y=\tau=2, p=1/2$) of up to $N=6000$ sites. The nominal $p-$value at which 
the clusters were grown was $p \approx 0.473$, the difference between this 
and the target value $p=1/2$ being made up by re-weighting. In this way
the variances of the observables were reduced by roughly two orders of 
magnitude, as compared to a straightforward Leath algorithm.

Results for the logarithm of the partition sum and for the gyration radius
will be shown later, together with results obtained for $y<2$.
They will be discussed in detail in the next section, here we just point
out that they are in perfect agreement with the expectations~\cite{Stauffer92}
\be
   Z^{\rm perc}_N(p=1/2) \sim N^{-5/91}\;,\quad R^2_N \sim N^{96/91}
\ee
and with simulations using the plain Leath algorithm.

More interesting are the average numbers of contacts and bonds, and 
the fluctuations thereof. Near the critical point, one has the scaling
ansatz~\cite{Stauffer92}
\be
   Z^{\rm perc}_N(p) \approx N^{-5/91} F((p-1/2)N^\sigma)     \label{perco-scale}
\ee
with $\sigma = 36/91$. This gives $\partial \ln Z^{\rm perc}_N/\partial p |_{p=1/2} \sim N^\sigma$
and $\partial^2 \ln Z^{\rm perc}_N/\partial p^2 |_{p=1/2} \sim N^{2\sigma}$.
On the other hand, using the defining equation (3), one obtains at $p=1/2$
\be
\left . { \frac{ \partial \ln Z^{\rm perc}_N}{\partial p}}\right |_{p=1/2} = {1\over 2} \langle b-s-k \rangle
\ee
and
\be
 \left .{\frac{\partial^2 \ln Z^{\rm perc}_N}{\partial p^2}} \right |_{p=1/2}= {1\over 4} \{{\rm Var}[b-s-k] - 
       \langle b+s+k \rangle\}\;.
\ee
Using this and eliminating $s$ in favour of $N$, one finds that 
\be
   \langle 3b+k \rangle = 4 N  + O(N^{\sigma}) \;,          \label{3bk}
\ee
while 
\be
   {\rm Var}[3b+k] = 2 \langle b\rangle + O(N^{2\sigma})\;.         \label{var}
\ee

Our estimates for the average numbers of bonds and contacts are 
\be
  \langle b\rangle = 1.1215(3)N + O(N^{\sigma}),\; \langle k\rangle = 0.6360(5)N + O(N^{\sigma}), 
\ee
in excellent agreement with Eq.~(\ref{3bk}).

Our results for the $2\times 2$ covariance matrix of the bond and contact numbers are 
shown in Fig.~2. More precisely, we show there the (co-)variances divided by $N$, 
\be
   C_{ij} = (\langle ij\rangle - \langle i \rangle\langle j  \rangle)/N.
\ee
From this figure we see first that $C_{bk}$ is negative for all $N$.
This is intuitively very plausible: The sum $b+k$ fluctuates less than $b$ and $k$ themselves. 
Moreover, $C_{bk}$ seems to scale as $N^\alpha$ with $\alpha \approx -0.09$, i.e. the 
covariance between bonds and contacts increases less fast than $N$. Secondly, we see that 
$C_{bb}$ tends to a constant for $N\to\infty$. A more precise analysis shows that the 
corrections to this are $\propto N^{-1/2}$, i.e. $C_{bb} \sim (1-{\rm const} /N^{1/2})$. 
Next, the variance of $3b+k$ seems to scale with a power of $N$ larger than 1, 
but this would contradict Eq.~(\ref{var}). Indeed, a more careful analysis shows that the 
data for ${\rm Var}[3b+k]$ are in perfect agreement with Eq.~(\ref{var}), and that the 
increase apparent of ${\rm Var}[3b+k]/N$ with $N$ is entirely due to the (predicted!) corrections to 
scaling. Finally, although it increases even faster with $N$, also ${\rm Var}[k]$ must 
ultimately scale $\sim N$, since it is just a linear combination of the previous (co-)variances.

\begin{figure}
  \begin{center}
    \psfig{file=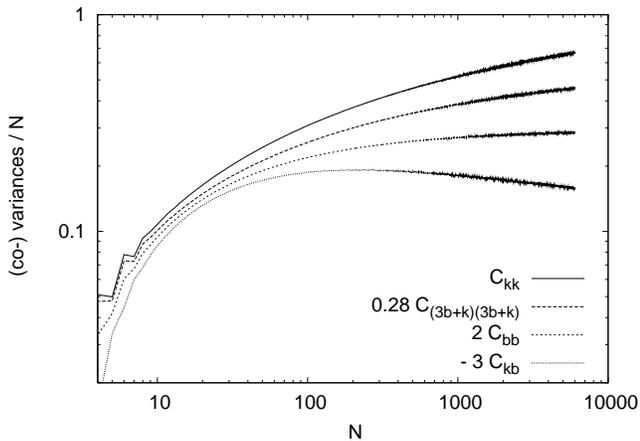,width=6.0cm,angle=270}
    \caption{Variances and covariance for bond and contact numbers at the critical percolation
     point, divided by $N$. Normalizations are chosen such that the curves do not overlap. Notice 
     that the covariance is negative for all $N$ and increases less fast than $N$. In contrast, 
     all three variances increase asymptotically $\sim N$, although with vastly different 
     finite size corrections.}
\label{fig2}
\end{center}
\end{figure}

We should stress again that all these results are as expected from the scaling theory for 
percolation (although they had not been derived or checked previously, to our knowledge~\cite{footnote}). 
They show that naive power law fits without guidance by the scaling theory would lead to 
$C_{kk} \sim N^\beta$ with $\beta > 0$ and thus to erroneous conclusions.

\section{The Region $y<2$: Collapsing Trees}

Let us first discuss the case $y=0$, and treat the general case $0<y<2$ later. In this 
case there are only the minimal number of bonds, $b=N-1$. To find the transition point
$\tau_c = \tau_c(0)$, we studied the scaling of the gyration radius and of the partition
sum. We also studied the specific heat (i.e., the variance of the number of contacts), 
but as in previous cases~\cite{animals04,adsorb} this gives much less precise estimates.

In Fig.~3 we plot the re-scaled squared gyration radii, $R^2_N/N^{2\nu}$, with a suitably 
chosen value of $\nu$, against $\ln N$. In Fig.~4 we show effective exponents $\theta_{\rm eff}$
defined by the triple ratios~\cite{DJ}
\be
   \theta_{\rm eff}(N,\tau) =  {7\ln Z_N - 6\ln Z_{N/3} - \ln Z_{5N} \over 6\ln 3 -\ln 5}.
                 \label{triple}
\ee
If we assume the scaling ansatz
\be
   Z_N(y,\tau) \sim \mu(y)^N N^{-\theta} G((\tau-\tau_c(y))N^\phi)\;,
\ee
these triple rations should tend to $\theta$ for $N\to\infty$, provided $\tau=\tau_c(y)$.

\begin{figure}
  \begin{center}
    \psfig{file=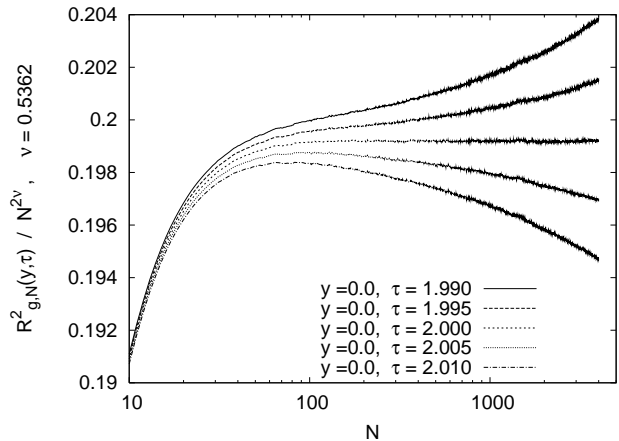,width=6.0cm,angle=270}
    \caption{$R^2_N/N^{2\nu}$ for $y=0$ and for five values of $\tau$ close to $\tau=2$,
     plotted against $\ln N$. The Flory exponent used for this plot was $\nu = 0.5362$.}
\label{fig3}
\end{center}
\end{figure}

\begin{figure}
  \begin{center}
    \psfig{file=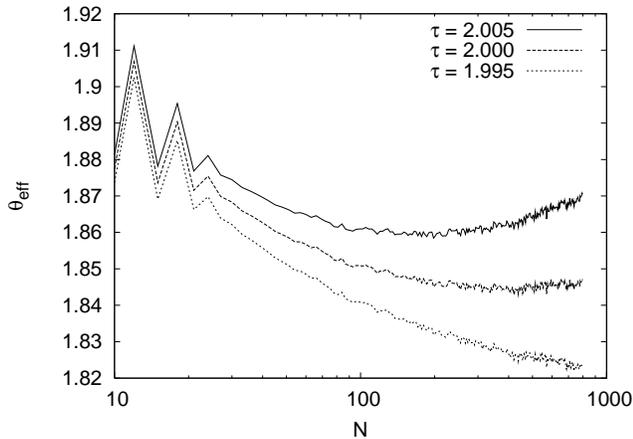,width=6.0cm,angle=270}
    \caption{Effective $\theta-$exponents, defined by Eq.~(\ref{triple}), plotted against
     $\ln N$. }
\label{fig4}
\end{center}
\end{figure}

From these plots we obtain $\tau_c(y=0) = 2.001 \pm 0.001$ (from Fig.~3) and 
$1.998\pm 0.002$ (from Fig.~4). We conjecture that the collapse transition 
occurs indeed exactly at $\tau_c=2$. This was already conjectured by previous
authors~\cite{Rensburg97}, although based on much more noisy data. Our best unbiased 
estimates of the critical exponents for collapsing trees are
\be
   \nu = 0.5359 \pm 0.0003, \quad \theta = 1.842 \pm 0.002.
\ee
If we accept the conjecture that $\tau_c=2$, then these estimates can be improved to 
\be
   \nu = 0.5362 \pm 0.0001, \quad \theta = 1.845 \pm 0.001.
\ee
These values are close to rationals with relatively small denominators, which might 
suggest that $\nu$ and $\theta$ are indeed simple rational numbers, $\nu=37/69
=0.53623...$ and $\theta = 59/32=1.84375$.  
Together with the estimated inverse critical fugacity and with results given below, 
these estimates are collected in Table~1.

 \begin{table}
 \begin{center}
 \caption{ Estimates of critical points $\tau_c(y)$, of critical exponents $\nu$
   and $\theta$, and of inverse critical fugacities $\mu(y)$.}
   \label{table1}
 \begin{ruledtabular}
 \begin{tabular}{lllll}
$y$  & $\tau_c(y)$  &  $\nu$       &  $\theta$  & $\mu(y)$ \\
\hline
0.0  &  2.0         &  0.5362(1)   &  1.845(1)   &   7.14893(2) \\
0.5  &  \quad "     &  \quad "     &  \quad "    &   7.3432(1) \\
1.0  &  \quad "     &  \quad "     &  \quad "    &   7.5485(1) \\
1.5  &  \quad "     &  \quad "     &  \quad "    &   7.7665(2) \\
2.0  &  2.0         &  0.52747...  &  2.05494... &   8.0  \\
2.25 &  1.994(2)    &  0.5230(12)  &  2.11(1)    &   8.1088(5)\\
2.5  &  1.970(4)    &  0.5223(8)   &  2.12(1)    &   8.1770(5)\\
3.0  &  1.880(3)    &  0.5220(4)   &  2.12(1)    &   8.2275(5)\\
3.5  &  1.733(7)    & $<0.524$     &  $< 2.05$   &   8.182(3) \\
3.75 &  1.642(6)    &  0.524(2)    &  $< 2.03$   &   8.109(5)\\
4.0  &  1.54(1)     &  0.524(2)    &  2.00(3)    &   8.035(10)\\
4.5  &  1.32(1)     &  0.520(3)    &  2.01(4)    &   7.88(2) \\
5.0  &  1.04(1)     &  0.521(3)    &  1.98(5)    &   7.63(3) \\
5.5  &  0.71(1)     &  0.522(4)    &  1.88(6)    &   7.32(4)
 \end{tabular}
 \end{ruledtabular}
 \end{center}
 \end{table}

Universality suggests that the same exponents should also describe the transition for all
$y<2$. But direct verification is less easy, since there are important corrections to scaling 
due to the cross-over from the percolation point. Making plots like Figs.~3 and 4 for $y>0$
would not give very clean results: Searching for cleanest power laws would give exponents
which depend on $y$, and would give mutually exclusive estimates $\tau_c(y)$ from the gyration 
radius and from the partition sum scaling. 

We therefore adopt a different strategy. Assuming that $\tau_c(y)=2$ for all $y\leq 2$ and 
that critical exponents are independent of $y$, we plotted in Fig.~5 $R^2_N/N^{2\nu}$ 
against $N$ for different values of $y$. Similarly, we plotted 
$\ln [Z_N N^\theta / \mu(y)^N]$ in Fig.~6, where $\mu(y)$ is carefully chosen such
as to take into account the dominant exponential increase of $\ln Z_N$ with $N$.
In both plots, the curves for percolation (i.e. the lower-most curves) become straight 
lines for $N\to\infty$, with slopes $2(\nu^{\rm perc}-\nu^{\rm tree})$ and $\theta^{\rm tree}-
\theta^{\rm perc}$ [notice that $\theta^{\rm perc}$ is usually called $\tau$ in 
the percolation literature~\cite{Stauffer92}].
The curves for trees ($y=0$) become horizontal. Finally, the curves for $0<y<2$ first follow 
the percolation curves and ultimately also 
become horizontal  for $N\to\infty$, but very slowly due to the slow cross-over.

\begin{figure}
  \begin{center}
    \psfig{file=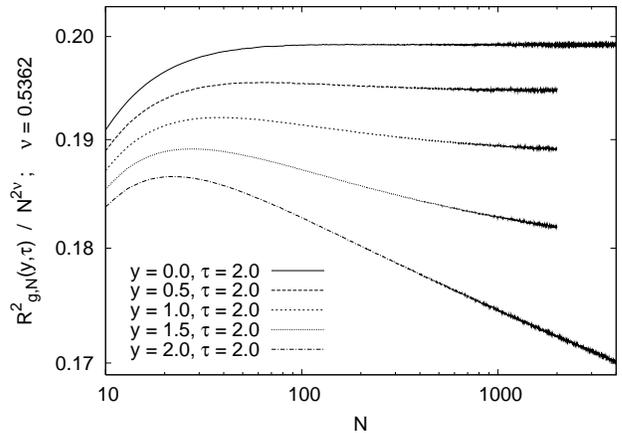,width=6.0cm,angle=270}
    \caption{Log-log plot of $R^2_N/N^{2\nu}$ for $\tau=2$ and for five values of $y$ 
     (0.0, 0.5, 1.0, 1.5, and 2.0).
     The Flory exponent used for this plot was $\nu = 0.5362$ as obtained from Fig.~3.}
\label{fig5}
\end{center}
\end{figure}

\begin{figure}
  \begin{center}
    \psfig{file=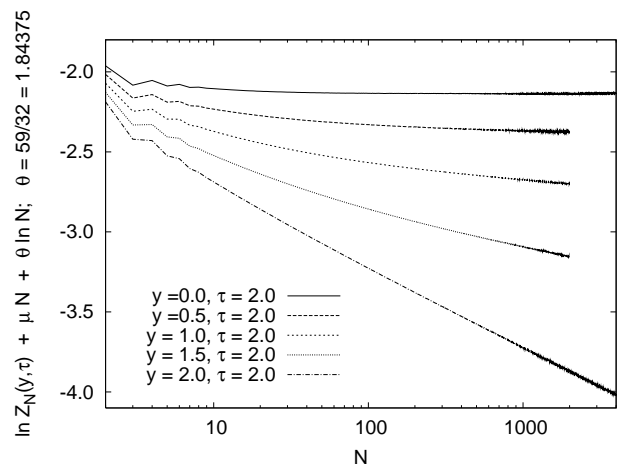,width=6.0cm,angle=270}
    \caption{Values of $\ln [Z_N N^\theta / \mu(y)^N]$ for $\tau=2$ and for the same five 
     values of $y$ as used in Fig.~5. The value of $\theta$ is that for $y=0$. The critical 
     inverse fugacities $\mu(y)$ were chosen such 
     that the curves show least bending for large $N$.}
\label{fig6}
\end{center}
\end{figure}

Although we cannot claim from Figs.~5 and 6 that this scenario is unique, they strongly 
suggest that there is indeed just a slow cross-over from critical percolation to collapsing 
trees, and that the collapse transition occurs for all $y<2$ exactly at $\tau=2$.

Let us finally discuss the fluctuations of the bond and contact numbers. The normalized 
variance of the number of contacts, $C_{kk}$, of collapsing trees (i.e. at $y=0$) is shown 
in Fig.~7 as a function of $\tau$
for various values of $N$. These data are in good agreement with the less precise results 
of previous simulations~\cite{Madras97,Rensburg99}. They verify that $\tau_c(0)\approx 2$,
but it would obviously be difficult to estimate from them $\tau_c(0)$ (as attempted in~\cite{Madras97,Rensburg99}) with a precision comparable to that obtained from Figs.~3
and 4.

\begin{figure}
  \begin{center}
    \psfig{file=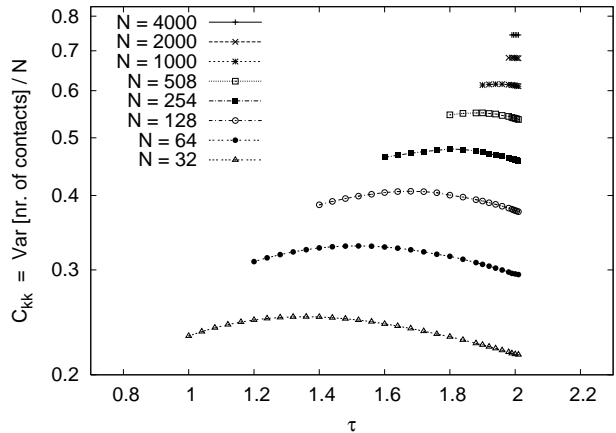,width=6.0cm,angle=270}
    \caption{Variances of contact numbers divided by $N$, plotted against $\tau$, for
     $y=0$ and for various values of $N$. }
\label{fig7}
\end{center}
\end{figure}

Fluctuations for different values of $y$ but precisely at $\tau=\tau_c(y)$ are shown in 
Figs.~8 to 10. In each figure, the normalized (co-)variance is plotted against $\ln N$.
For $y\leq 2$ the curves correspond to $\tau=2$, while
they represent our best estimates of the collapse transition for $y>2$.

\begin{figure}
  \begin{center}
    \psfig{file=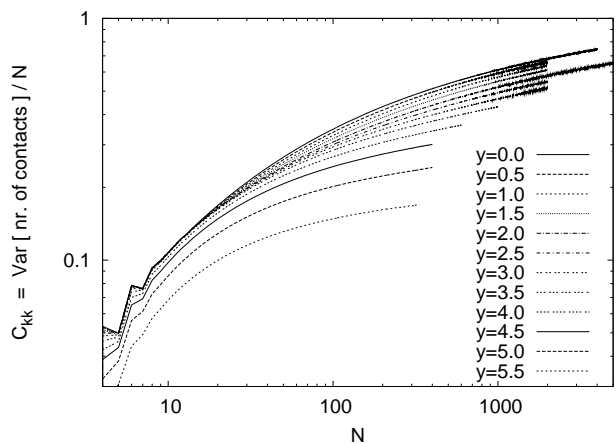,width=6.0cm,angle=270}
    \caption{Variances of contact numbers divided by $N$, plotted against $\ln N$. For 
     each $y$ the curve is for our best estimate of $\tau_c(y)$.}
\label{fig8}
\end{center}
\end{figure}

\begin{figure}
  \begin{center}
    \psfig{file=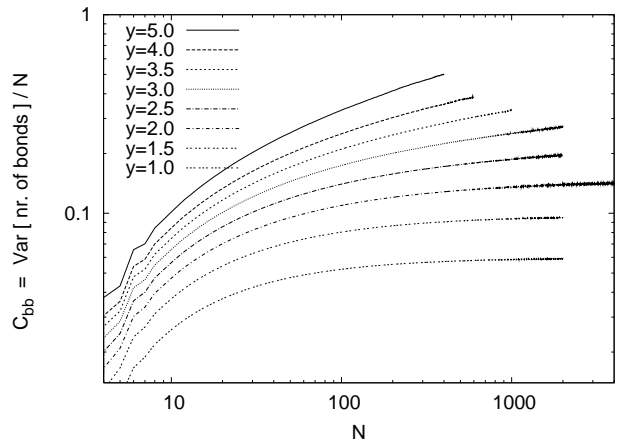,width=6.0cm,angle=270}
    \caption{Same as Fig.~8, but for number of bonds.}
\label{fig9}
\end{center}
\end{figure}

\begin{figure}
  \begin{center}
    \psfig{file=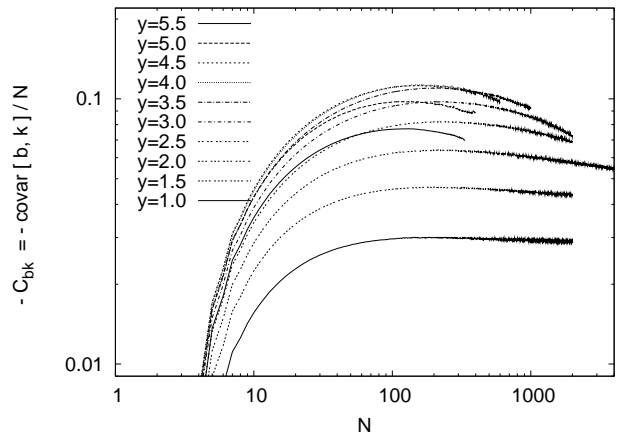,width=6.0cm,angle=270}
    \caption{Similar to Figs.~8 and 9, but for the covariances between bond and contact numbers.
     Since these covariances are negative for all $y$ and $N$, we plot actually $-C_{bk}$.}
\label{fig10}
\end{center}
\end{figure}

As expected, the fluctuations of $b$ are very small and tend to zero as $y\to 0$. They scale 
$\sim N$. The bond-contact covariances seem to increase slower than $N$ for all $y>0$, although
the decrease of $|C_{bk}|$ with increasing $N$ becomes weaker for small $y$. Of most 
interest (for the present case $y<2$) is the scaling of $C_{kk}$, since it is the 
contacts which should drive the collapse transition. 

Obviously $C_{kk}$ increases with $N$. Naive fits (e.g. least square) would give power 
laws $C_{kk} \sim N^{2\phi-1}$ with $\phi \approx 0.6$ to $0.65$. This would agree with
previous estimates~\cite{Seno94,Flesia94,Madras97,Rensburg97,Rensburg99}, but we should 
be extremely careful with accepting such a fit. The reason 
is that there is no qualitative difference between the cases $y<2$ and $y=2$, and for the 
latter we had seen in Sec.~3 that $C_{kk} \to {\rm const}$ for $N\to\infty$. Thus we propose 
that $C_{kk} \to {\rm const}$, i.e. $\phi=1/2$, also for $y<2$. If we would have $\phi>1/2$, 
then there should be either a clear cross-over (which is not seen in Fig.~10), or $\phi$
would have to be non-universal.

\section{The region $y>2$}

For $y>2$ the clusters are richer in bonds, and less rich in contacts. The collapse 
transition was again located by requiring both the gyration radius and the partition sum 
to scale. But this time it is much more difficult to estimate corrections to scaling. The 
main reason is that our algorithm deteriorates very rapidly when $y$ becomes large. While 
it is still efficient for $y\leq 3.5$, it becomes virtually useless for $y\geq 5.5$. According 
to~\cite{Derrida-H}, the end point of the transition line is at $y=y_{\rm DH}=6.48\pm 0.12$.
At this point, the present algorithm was unable to give reasonable statistics of clusters 
with $N=200$. The algorithm based on site animals discussed in~\cite{animals04} does a 
bit better and allowed us to obtain a good sample of clusters with $N=300$. But even 
for this sample errors are quite large, and extrapolation to $N\to\infty$ obviously 
becomes very difficult. We thus do not present any data for the Derrida-Herrmann model
($\tau=0,y=y_{\rm DH}$), but we just state that our simulations were in complete agreement 
with the results of~\cite{Derrida-H}.

We therefore must obtain estimates of the critical parameters from the region $2 <y < 5$, 
although there could again be important cross-over contributions from the percolation 
point.

Typical results for $y=3.0$ are shown in Figs.~11 and 12. In Fig.~11 we see that
gyration radii suggest the transition to be at $\tau_c(y=3.0) = 1.878\pm 0.003$, with
$\nu = 0.5222 \pm 0.0005$. Effective $\theta-$exponents defined by Eq.~(\ref{triple}) are 
shown in Fig.~12. They corroborate the determination of $\tau_c$, more precisely they 
suggest $\tau_c(y=3.0) = 1.884\pm 0.004$. Using as a compromise $\tau_c(y=3.0) = 1.880\pm 0.003$,
we obtain as our best estimates of the critical exponents $\nu = 0.5220 \pm 0.0004$ and
$\theta = 2.12\pm 0.01$.

\begin{figure}
  \begin{center}
    \psfig{file=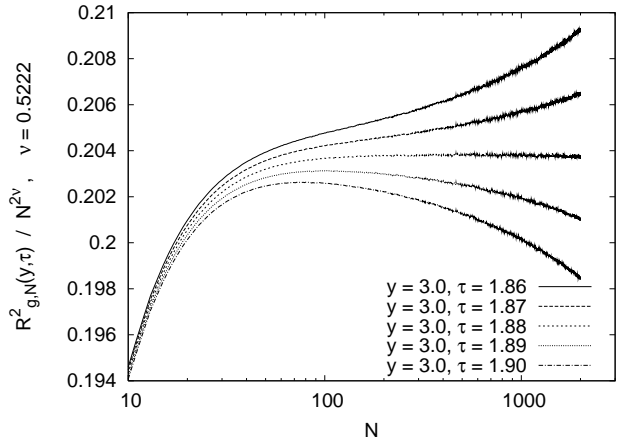,width=6.0cm,angle=270}
    \caption{Average squared gyration radii for $y=3.0$, divided by $N^{1.0444}$ and plotted
     against $\ln N$. }
\label{fig11}
\end{center}
\end{figure}

\begin{figure}
  \begin{center}
    \psfig{file=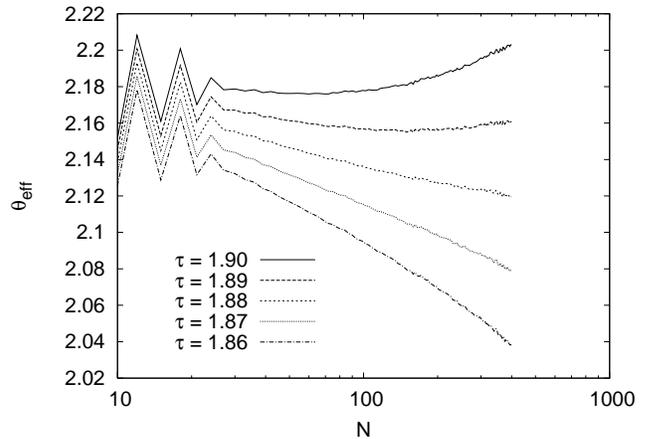,width=6.0cm,angle=270}
    \caption{Effective exponents $\theta_{\rm eff}$ at $y=3.0$, plotted against $\ln N$. }
\label{fig12}
\end{center}
\end{figure}

Estimates of $\nu$ and $\theta$ compatible with these were obtained for $y=2.5$ and $y=2.25$, 
with $\tau_c(y=2.5)=1.970\pm 0.004$ and $\tau_c(y=2.25)=1.994\pm 0.002$. But we encountered 
problems when going to $3.5 \leq y \leq 4$. Results analogous to Figs.~11 and 12, but for
$y=3.75$, are shown in Figs.~13 and 14. This time, there is no value of $\tau$ at which 
$R_N^2$ shows a pure power law for large $N$ (say, $N>100$). For small $\tau$ the curves 
bend upward (the clusters are extended), while for $\tau\geq 1.630$ they bend down for very
large $N$, showing that the collapse sets first in for very large clusters only. Moreover, 
taking the curve in Fig.~13 for $\tau = 1.625$ (which seems to become straight for very large $N$) 
to estimate $\nu$, we would obtain $\nu=0.527$, which is definitely much larger than the
estimates obtained from $2.5 \leq y \leq 3.0$.

\begin{figure}
  \begin{center}
    \psfig{file=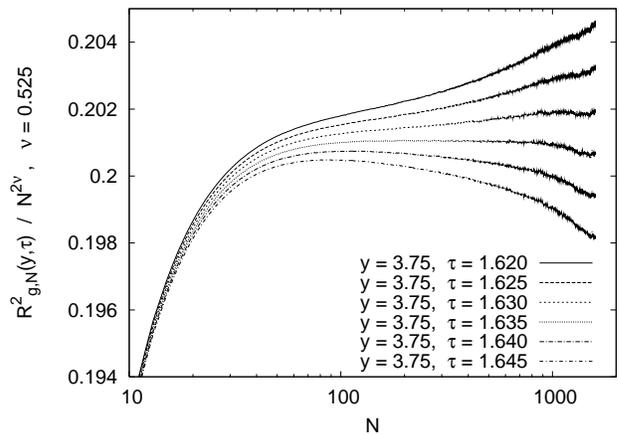,width=6.0cm,angle=270}
    \caption{Average squared gyration radii for $y=3.75$, divided by $N^{1.05}$ and plotted
     against $\ln N$. }
\label{fig13}
\end{center}
\end{figure}

\begin{figure}
  \begin{center}
    \psfig{file=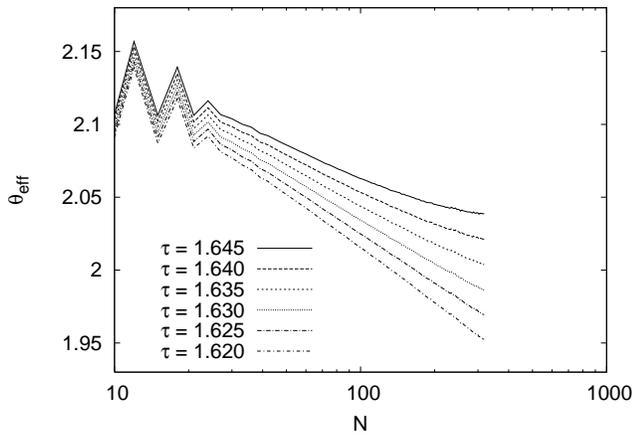,width=6.0cm,angle=270}
    \caption{Effective exponents $\theta_{\rm eff}$ at $y=3.75$, plotted against $\ln N$. }
\label{fig14}
\end{center}
\end{figure}

Similarly, also the partition sum shows late scaling, 
with all curves in Fig.~14 decreasing. Only the curves for $\tau \geq 1.64$ 
shows some tendency to level off for very large $N$. If this is taken as an indication that 
scaling should set in similarly late also for the gyration radii, then indeed the lowest 
curve in Fig.~13 should be the critical one, and its slope for $N>1000$ should give 
the correct estimate of $\nu$: $\nu \approx 0.520$. But all this looks very unconvincing. 
In addition, the value of $\theta$ suggested from Fig.~14 is much smaller 
than that from Fig.~12, any reasonable estimate based on Fig.~14 being $\theta \leq 2.06$.

Similar results were obtained for $y=3.5$ and 4.0. In all these cases the estimated
value of $\theta$ is definitely smaller than the estimate obtained from $y=2.5$ and 3.0. 
Our best estimates from the region  $y\geq 3.5$ are $\theta < 2.0, \nu < 0.515$.
Notice that this is not easily understood from a cross-over from percolation, since there
$\theta = 187/91 = 2.054945$. Why should the estimates first move away from the percolation
value as we move away from the percolation point, just to come back to it later? 

Notice that these difficulties, serious as they are for the estimation of the critical exponents,
have very little effect on the estimation of $\tau_c(y)$. In spite of them, our estimates
(shown in Fig.~1) have errors less than $0.02$. 

They have also very little influence on 
the estimates of the bond and contact variances. These depend rather weakly on $\tau$ 
(thus they would not be very useful for determining the collapse transition curve, as 
proposed in all previous works). The errors in the data shown in Figs.~8 to 10 are comparable
to the thickness of the lines. As expected, $C_{kk}$ decreases as $y$ gets larger, while 
$C_{bb}$ increases. Thus we can safely say that 
\be
   C_{kk} \to {\rm const} \quad {\rm for}\quad N\to\infty\;,
\ee
since this was already shown for $y=2$. The scaling of $C_{bb}$ is much more subtle. 
Previous analyses, starting with the 
seminal work of~\cite{Derrida-H}, obtained $C_{bb} \sim N^{2\phi-1}$ with $\phi \approx 0.6$ 
to 0.65. Just making least square fits to the curves in Fig.~9 for the largest values of $y$
would give the same estimate. But we think that this is misleading, as it was also for
$y\leq 2$. First of all, we would expect some sort of cross-over in Fig.~9, which is not 
seen: The transition from $y=2$ to $y>2$ is perfectly smooth. Secondly, we should expect 
$C_{kk}$ and $C_{bb}$ to scale in the same way. These two variances control the divergence 
of $Z_N(y,\tau)$ as we cross the transition curve vertically and horizontally. In both cases 
one crosses the transition curve {\it transversally}, and along each transverse line the 
divergence should be the same. Therefore we suggest that $C_{bb}\sim {\rm const}$, i.e. $\phi=1/2$.

\section{Collapsed animals: One or Two Phases?}

According to~\cite{Flesia92,Flesia94,Peard95} there are indeed two collapsed phases (one 
rich in contacts, the other rich in bonds). The transition line between them bifurcates 
off from the collapse curve not at $y=y_c = 2$, but at $y^* \approx 4$ (see the figure 
shown in~\cite{Henkel96}).  If this is basically true, but with $y^*\approx 3.2$, this 
would easily explain our results discussed in the previous section: The point $(y^*,\tau^*)$ 
would separate an intermediate part $2 < y \leq y^*$ of the collapse curve which is still 
basically contact driven from the bond driven part $y>y^*$. The only unusual feature would 
be that the fixed point $(y^*,\tau^*)$ has to be attractive when approached from the left 
along the collapse transition curve, while it has to be repulsive on the right. Such a 
behaviour is not typically expected from the RG group, although we see no reason why it 
should be forbidden (it requires that the linearized RG flow near the fixed point is zero, 
and the flow is dominated by quadratic terms).

For a direct test of this scenario, one has to simulate deep in the collapsed region and verify 
that there is a transition between two different collapsed phases. To search for such a transition, 
we have to modify our strategy. We cannot use the gyration radius as an indicator (because 
$R_N^2 \sim N$ in both phases), and we cannot use the partition sum either. We checked that 
the free energy of collapsed animals has a surface term, $\ln Z_N = N \ln\mu - {\rm const} N^{1/2} + 
\ldots$, as one expects for any 2-d compact object (a similar term is well known for droplets, 
and was shown to exist also for collapsed unbranched polymers~\cite{owczarek,hsu}). Even if 
there is in addition a term $\sim \ln N$ in the free energy, there would be just too many 
unknown parameters to pin it down precisely and to use this for location the transition.

\begin{figure}
  \begin{center}
    \psfig{file=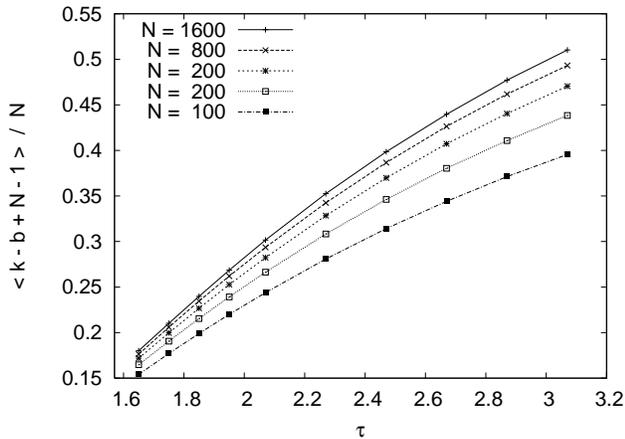,width=6.0cm,angle=270}
    \caption{Normalized average differences $(k-b')/N$ at $y=3.75$ plotted against $\tau$, for 
     $N=100,200,400,800,$ and $1600$ (from bottom to top). Here, $b' = b - N+1$ is the number 
     of bonds exceeding the minimal number needed for the cluster to be connected. Error 
     bars are much smaller than the size of the symbols.}
\label{aver_kb}
\end{center}
\end{figure}

Thus we have to look directly to the order parameter, which is the difference $b-k$, and its 
fluctuations~\cite{footnote2}. In Fig.~\ref{aver_kb} we show the normalized average difference
$\langle k-b\rangle / N$ between the numbers of contacts and bonds against $\tau$, for several 
values of $N$. More precisely, we used $b-N+1$ instead of $b$ in this figure, which is the 
number of bonds exceeding the minimal number. For this figure, $y$ was kept fixed at $y=3.75$, 
i.e. we scan Fig.~1 along 
a vertical line which starts at the collapse transition line at a $\tau$ which the previous 
section suggested to be above $\tau^*$. We see that this difference increases with $\tau$, 
indicating that we move indeed from a situation rich in bonds to a system rich in contacts. 
But this increase is so smooth that this cannot be used as an argument in favour of a 
thermodynamic phase transition. Instead, Fig.~\ref{aver_kb} could suggest that there is no 
such transition. But we should be aware that an analogous search for the $\theta$ 
collapse in ordinary (unbranched) polymers, based on the number of contacts, would also 
fail, because this number varies rather smoothly with temperature~\cite{GH95}. 
Thus we would need much larger systems to obtain a firm conclusion from a plot like 
In Fig.~\ref{aver_kb}.

\begin{figure}
  \begin{center}
    \psfig{file=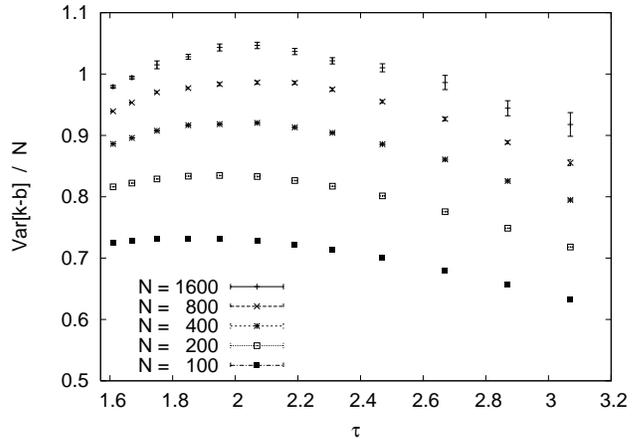,width=6.0cm,angle=270}
    \caption{Analogous to Fig.~\ref{aver_kb}, but showing the normalized variances
     $C_{k-b,k-b}$.}
\label{var_kb}
\end{center}
\end{figure}

Things change if we look at the corresponding variances, see Fig.~\ref{var_kb}. If there is 
a phase transition, we expect a peak which becomes higher as $N$ increases. What we see is 
not quite a sharp peak (it is rather a broad bump), but its hight definitely increases with
$N$. Also, its maximum shifts with increasing $N$ to higher values of $\tau$, indicating that 
it is not related to the collapse transition. Notice that the existence of this bump and its
increase with $N$ is definitely not due to any statistical fluctuations. 

It is not clear whether the data shown in Figs.~\ref{aver_kb} and~\ref{var_kb} taken 
by themselves speak more for or against are phase transition between two collapsed phases.
They certainly do not exclude this possibility. And taken together with the anomaly found 
in the previous section, we believe that such a transition is the most natural interpretation.
In this case, a naive (least square, say) fit to an ansatz $C_{k-b,k-b} \sim N^{2\phi-1}$ 
would suggest $\phi\approx 0.54$, but there is the same kind of curvature in plots of 
$C_{k-b,k-b}$ versus $\ln N$ as in Figs.~\ref{fig8} and ~\ref{fig9}, and a more 
careful extrapolation suggests again that $\phi$ is close to $1/2$, more precisely 
$\phi = 0.52 \pm 0.02$. Since the shift of the maximum to larger $\tau$ with increasing 
$N$ seems to stop for $N>400$, we locate the transition near the maximum observed for
$N=1600$, $\tau \approx 2.1\pm 0.2$. Because 
these simulations were very costly in terms of CPU time, we have not made similar searches
at different values of $y$, and the location of the transition line indicated in Fig.~1
should only be taken as a very rough guess.

We should however point that it seems hard to come by with a theoretical argument for 
such a transition. Usually, a phase transition requires some sort of cooperativity, i.e.
some mechanism leading to positive correlations. In the ferromagnetic Ising model, e.g., 
this is the spin-spin interaction which tends to make neighbouring spins parallel. In the 
antiferromangetic Ising model on a square lattice, the spin-spin interaction favours
parallel next-nearest neighbours. In the present case we see no such interactions. The 
only source for correlations is the fact that the bonds must make the cluster connected, 
i.e. they cannot all be in one part of the cluster, leaving the rest of the cluster to 
contacts. But this should give rise to negative correlations, and it is not clear how 
it can lead to a phase transition.

We should also mention that the individual bond and contact variances variances show no 
bump or peak when going into the collapsed region, but fall off continuously with $\tau$.
The bump seen in the variance of $k-b$ is entirely due to an increase of the absolute 
value of the covariance. Indeed, as we move off the collapse curve into the collapsed phase,
the scaling of $C_{kb}$ changes from decreasing with $N$ to increasing with it, and 
continues to increase with $N$ even for very large $\tau$.

\section{Conclusions}

We have presented extensive Monte Carlo simulations of collapsing lattice animals and collapsing
lattice trees. We used a novel algorithm which should be most efficient near the (bond)
percolation point, and it indeed was. At the percolation point it was even more efficient
than the straightforward Leath algorithm. It was also very efficient along the entire 
transition line for $y<2$, including the point $y=0$ which describes the collapse of weakly 
embeddable trees. For the latter we conjectured exact values for all critical exponents.

Our simulations encountered severe problems for the bond-induced collapse at $y>2$. For 
very large $y$, near the point studied first by Derrida and Herrmann, it essentially 
breaks down. Thus we were able to verify their analysis, but we were not able to 
improve on it.

For moderately large $y$, i.e. for $3.5 \leq y \leq 4.0$, our algorithm is still efficient
enough to generate high statistics samples of rather large clusters ($N \approx 1000$).
But we found there very large corrections to scaling which prevent us from extracting
precise values of the critical exponents, and we can only give upper bounds for 
$\nu$ and $\theta$: $\nu < 0.515, \theta < 2.0$. These corrections seem to be absent for even 
smaller $y$ (i.e. $2.5 \leq y \leq 3.0$), but we cannot of course exclude that they 
would show up if we would go to even larger clusters. Assuming this not to happen, 
we can then propose $\nu = 0.5220 \pm 0.0004, \phi = 0.5$ and $\theta = 2.12 \pm 0.01$ 
along the bond-induced collapse transition, while $\nu = 37/69 \approx 0.53623, \phi = 0.5$ 
and $\theta = 59/32 = 1.84375$ for collapsing trees. Together with the exact results
$\nu = 48/91 \approx 0.52747, \phi = 1/2$, and $\theta = 187/91 \approx 2.054945$ for critical 
percolation, these values show that there are indeed (at least) three different 
universality classes. We cannot rule out that the anomalies observed at $y\approx 3.75$
hint at yet another universality class. 

We found somewhat weaker but still statistically significant evidence for a 
transition between two collapsed phases, one contact-rich and the other bond-rich. 
This line branches off from the collapse line at $y^* = 3.2\pm 0.2$, which is 
clearly larger than the value $y=2$ for critical percolation, but is substantially
smaller than the estimate $y^*\approx 4$ given in~\cite{Flesia92,Flesia94,Peard95}. The 
cross-over exponent at this transition is again $\phi \approx 0.5$. Other critical 
exponents were not measured for this transition.

We should finally comment on our claim that $\phi=1/2$ for critical percolation
(and indeed along the entire collapse line and for the transition between the two 
collapsed phases). This 
is based on the definition of $\phi$ in terms of bond and/or contact fluctuations. 
It is also true for site percolation, if we look at fluctuations in the number of 
non-wetted surface sites. It should not be confused with the fact that the cross-over 
exponent in a scaling ansatz, Eq.~(\ref{perco-scale}), usually called $\sigma$ in 
the percolation literature, is smaller than 1/2. From this scaling ansatz it follows
surprisingly that the variances of bond and contact numbers must scale $\sim N$. 
Our claim that $\phi=1/2$ is entirely based on this observation, and is made 
in spite of the fact that naive fits would give $\phi > 1/2$. This should again 
be a warning that power law fits not guided by a solid theory can be very misleading.

Acknowledgement: It is a pleasure to thank Walter Nadler for numerous discussions 
and for critically reading the manuscript.


\begin{thebibliography}{99}
\bibitem{footnote} For critical site percolation one has $\langle s\rangle = (1-p_c)N/p_c + O(N^\sigma)$
   and ${\rm Var}[s] = (1-p_c)N/p_c^2 + O(N^{2\sigma})$, where $s$ is the number of non-wetted perimeter
   sites and $p_c = 0.5927...$. The only paper dealing with fluctuations of bonds, contacts, or 
   perimeter sites in percolation we are aware of is~\cite{Borstnik}, but these authors were 
   unaware of the fact that these fluctuations could be derived from accepted scaling ans\"atz.
\bibitem{footnote2} Define a spin variable $s_i = 1$ if the edge $i$ between two occupied adjacent
   lattice sites is occupied by a bond, and $s_i = -1$ if it is occupied by a contact.
   We could then have also looked at spatial spin-spin correlations, but we shall not discuss
   this further in the present paper.
\end{thebibliography}
\end{document}